\def\DraftSize{
 \documentstyle{article}
 \textwidth 17cm
 \textheight 24cm
 \topmargin -40pt
 \oddsidemargin -20pt
 \evensidemargin -20pt
 \renewcommand{\baselinestretch}{0.95}
 }

\def\FullSize{
 \documentstyle[12pt]{article}
 \textwidth 16.5cm
 \textheight 23cm
 \oddsidemargin -0.2cm
 \evensidemargin -0.2cm
 \topmargin -1cm
 \parindent=3em
 \renewcommand{\baselinestretch}{1.4}
}

\FullSize

\def\beq{\begin{equation}}
\def\eeq{\end{equation}}
\def\bea{\begin{eqnarray}}

\def\eea{\end{eqnarray}}
\def\ds{\displaystyle}

\def\ssz{\scriptsize}

\def\ni{\noindent}

\def\req#1{(\ref{#1})}

\def\rep{\mbox{Re}\ }

\def\Zed{\mbox{\bf Z}}

\def\barc{\begin{array}{c}}
\def\ear{\end{array}}
\def\text#1{\mbox{\ssz{#1}}}
\def\percent{\%\ }

\renewcommand{\baselinestretch}{1.4}

\begin{document}

\ni{\Large
{\bf Regularized Casimir energy for an infinite dielectric cylinder subject to
light-velocity conservation}} \\
\vspace*{5mm}
\ni Israel Klich$^a$\footnote{E-mail: klich@tx.technion.ac.il} and
August Romeo$^b$\footnote{E-mail: romeo@ieec.fcr.es} \\
\ni a. Department of Applied Mathematics and Physics, Technion, 32000 Haifa, Israel \\
\ni b. CSIC research unit at IEEC (Institute for Space Studies of Catalonia), c. Gran Capit{\`a} 2-4, 08034 Barcelona
\vspace*{5mm}

\ni{\bf Abstract.}
The Casimir energy of a dilute dielectric cylinder,
with the same light-velocity as in its surrounding medium,
is evaluated exactly to first order in
$\ds\xi^2=\left( \varepsilon_1-\varepsilon_2\over \varepsilon_1+\varepsilon_2 \right)^2$
(where $\varepsilon_1, \varepsilon_2$ are the dielectric constants of the
cylinder and of its environment),
and numerically to higher orders in $\xi^2$.
The first part is carried out using
addition formulas for Bessel functions,
and no Debye expansions are required.

\ni{\small PACS: 03.70.+k, 12.20.-m, 42.50.Lc}

\section{Introduction}
Zero-point fluctuations of quantum fields give rise to forces, which are
regarded as manifestations of
the Casimir effect (for reviews, see e.g. refs.\cite{PMG}). From the
theoretical viewpoint,
one of the most daunting aspects of the evaluation of Casimir energies,
even for highly symmetrical boundaries,
is its sheer difficulty.
Many
mathematical methods has been developed,
but even the simplest ones demand considerable efforts.

At the core of several of these techniques one finds
uniform asymptotic expansions
---also called {\it Debye} expansions--- of Bessel or Ricatti-Bessel functions
appearing in integrals over momentum-like variables. This fruitful
method was used as early as ---at least--- the time of
ref.\cite{MDS78}, and has been repeatedly revisited in a huge
number of articles, often in the framework of other regularization
schemes (see e.g. ref.\cite{LRap} and refs. therein). However
reliable, the whole Debye expansion technique is a time-consuming
process and the search for computational alternatives might be of
interest \cite{Kl}. This is, precisely, one of the motivations of
the present letter. Our purpose is to take further the
exploitation of summation theorems for Bessel functions, started
in ref. \cite{Kl} for cases with spherical surfaces, and apply it
to a problem with a cylindrical boundary.

We are considering
a material cylinder of radius $a$, infinitely long, placed along the $z$-axis,
with permitivitty and permeability
$\varepsilon_1, \mu_1$,
surrounded by a medium with
permitivitty and permeability
$\varepsilon_2, \mu_2$.
For such surfaces,
a special situation is the case where
the light velocities in
both media ---interior (1) and exterior (2)--- are the same, i.e.,
\beq
\varepsilon_1 \mu_1 = \varepsilon_2 \mu_2 \equiv c^{-2},
\label{ccons}
\eeq
where $c$ is the common light-velocity.
Since any variation in $\varepsilon$ affects $\mu$, this
is called {\it dielectric-diamagnetic} case, as opposed to the
{\it purely dielectric} one, in which $\mu_1=\mu_2=1$ but the velocity
has to change.
Dielectric-diamagnetic conditions are often
desirable
as they cause the frequency equations to simplify and some
divergences to cancel out.
In a QCD context,
$\varepsilon$ and $\mu$ refer to colour permitivitty and permeability
(see \cite{BKBN} and refs. therein).
Illustrations of the dependence of the interquark potential on the boundary
conditions for a string model have been provided in ref.\cite{KLN}.

In refs. \cite{MNN}, \cite{LNB} and \cite{NP} the regularized Casimir energy
per lateral unit-length for an infinite dielectric-diamagnetic cylinder has been
studied.
Up to the order of
\beq
\xi^2=\left( \varepsilon_1-\varepsilon_2\over \varepsilon_1+\varepsilon_2 \right)^2,
\label{defxi2}
\eeq
the energy has been shown to vanish within all
the tested degrees of numerical accuracy.
The next contribution,
which is of the order of $\xi^4$,
has been found
---to our knowledge, for the first time--- in ref.\cite{NP}.

In our minds,
medium 2 will be pure vacuum
and medium 1 a very tenuous dielectric,
which means
$\varepsilon_2=\mu_2=1$ and $\varepsilon_1-1 \ll 1$.
As a result, the $\xi^2$ parameter, defined by eq.\req{defxi2},
is a small number.
According to ref.\cite{MNN}, the eigenfrequencies $\omega$
coming from the Maxwell equations for this problem
are given by the zeros of some equations of the type
$\ds f_n(k_z,\omega,a)=0$, $n\in\Zed$
---eqs.(2.3)-(2.5) in ref.\cite{MNN}---.
Further, in cases where the relation \req{ccons}
holds,
$f_n$ takes the form
\beq
f_n(k_z, \omega, a )
=\ds -a^2 c^{-2} \lambda^6 { (\varepsilon_1+\varepsilon_2)^2 \over
4 \, \varepsilon_1 \, \varepsilon_2  } \,
\left[
\xi^2 {\cal P}_n^2(\lambda a)
+{4 \over \pi^2 (\lambda a)^2}
\right] ,
\hspace*{0.5cm} {\cal P}_n(x)\equiv\left( J_n H_n \right)'(x),
\label{defcalPW}
\eeq
where $\xi^2$ is given by \req{defxi2},
$J_n$, $H_n$ are Bessel and Hankel functions,
and
\beq
\omega=\ds c \, \sqrt{ \lambda^2 + k_z^2 }.
\eeq
Every $\lambda$ belongs to the eigenfrequency set of the
projected two-dimensional problem ---say $\Lambda$---,
while $-\infty < k_z < \infty$, i.e., the values of $k_z$ are
continuous without any restriction.
Before regularizing, the Casimir energy per unit-length (${\cal E}_C$)
is given by the mode sum
\beq
\ds
{\cal E}_C
= \ds{1 \over 2} \, \hbar \,
\sum_{n,m} \,
\int {dk_z \over 2\pi} \,
\omega_{n,m,k_z},
\hspace*{1.5cm}
\ds\omega_{n,m,k_z} = \ds c \, \sqrt{ \lambda_{n,m}^2 + k_z^2 }.
\label{decom}
\eeq
The $n$-index is the angular momentum number, while $m$ describes the
remaining degree of freedom,
i.e., labels the different $\lambda$-values at a given $n$.

The present work is organized as follows.
In sec. 2 we follow ref. \cite{MNN} and
evaluate the energy density
$g^{(2)}$ (in momentum space) up to the order of $\xi^2$,
by a modified Bessel function summation theorem,
and resorting to the properties of Meijer $G$ functions.
Then, we show that the integration of $g^{(2)}$ yields a vanishing result.
Sec. 3 is devoted to an alternative approach based on
a zeta function prescription for the initial mode sum
%(the method called {\it complete} zeta function in refs.\cite{LRap,GR}, among others).
like in refs.\cite{LRap,GR}.
Apart from proving to be easier, this technique
paves the way to the numerical calculation of
higher order contributions.
Our conclusions are given in sec. 4.

\section{Density method \label{secden}}

%We begin by briefly reviewing the derivation proposed in ref.
%\cite{MNN} for the Casimir energy.
We begin by reviewing the procedure used in ref. \cite{MNN} and
obtaining an expression for the Casimir energy.
%\req{decom}.
The mode sum is
first represented, as usual in these cases, by a contour integral
\begin{equation}
\label{contourlambda} E_C=-\frac{\hbar c}{2}\int_{-\infty}^\infty
\frac{dk_z}{2\pi}\sum_{n= -\infty}^{\infty} \frac{1}{2\pi
i}\frac{1}{2}\oint_{C}\sqrt{\lambda^2+k^2_z} \, d_\lambda
\ln\left[ \frac{f_n(k_z, \omega, a )} {f_{n,as}(k_z, \omega)} \right] \,{,}
\label{contoursum}
\end{equation}
where the integration contour $C$ consists of a straight line
parallel to, and just to the right of, the imaginary axis,
$(-i\infty,+i\infty)$ closed by a semicircle of an infinitely
large radius in the right half-plane.
%Following ref. \cite{MNN},
The branch line of the function
$\varphi(\lambda)=\sqrt{\lambda^2 +k_z^2}$ is chosen to run between
$-i|k_z|$ and $i|k_z|$ on the imaginary axis. In terms
of $y= \mbox{Im} \, \lambda$  we have
\begin{equation}
\varphi (iy) = \left \{
\begin{array}{lcl}
i\sqrt{y^2-k_z^2}, && y>k_z,\\ \pm\sqrt{k_z^2-y^2}, && |y|<k_z,\\
-i\sqrt{y^2-k_z^2}, && y<-k_z .
\end{array}
\right .
\end{equation}
Noting that the argument of the logarithm is an even function
%with respect to
of
$iy$,
%\footnote{This follows easily since this analytic
%function takes real values on the imaginary axis, thus $\Delta_n
%^{\text{TE,TM}}(iy)=\overline{ \Delta_n(-iy)
%^{\text{TE,TM}}}=\Delta_n ^{\text{TE,TM}}(-iy)$}
%, the expression
\req{contoursum} reduces to
\begin{equation}
E_C=-\frac{\hbar c}{2\pi^2}\sum_{n=-\infty}^{\infty}\int_0^\infty dk_z
\int_{k_z}^\infty \sqrt{y^2-k_z^2}\,d_y\,\ln\left[ 1-{{\xi
}^2}{{(y{{\partial }_y}({I_n}(a y){{{K}}_n}(a y)))}^2}
\right]\, ,
\label{EneExp}
\end{equation}
where we have expressed \req{defcalPW} explicitly on the imaginary
axis. Integrating with respect to $k_z$ we obtain the Casimir
energy per unit length (${\cal E_C}$) as an integral over $y$, namely
\begin{equation}
{\cal E_C}=\frac{\hbar c}{4\pi} \sum _{n=-\infty}^{\infty
}\int_{0}^{\infty } dy \, {y^2}{{\partial }_y} \ln \left[ 1-{{\xi
}^2}{{(y{{\partial }_y}({I_n}(a y){{{K}}_n}(a y)))}^2}
\right]\,{.} \label{ECLfromMNN}
\end{equation}
In the following subsections, we use the approach of ref. \cite{Kl}
to evaluate this expression.
\subsection{Density calculation to the order of $\xi^2$}
Having integrated $k_z$ out, the integrand in expression
\req{ECLfromMNN} may be interpreted as the density of Casimir
energy with respect to the parameter $y=i\lambda$. This density
may be evaluated by expanding in terms of $\xi^{2}$, the first
contribution being
\begin{equation}
g^{(2)}(y)\equiv \frac{\hbar c{\xi}^2}{4\pi }\sum_{n=-\infty}^{\infty }
d y \, {y^2}{{\partial}_y} {{\left[ y{{\partial }_y}(I_n(a y)K_n(a
y)) \right] }^2}. \label{E2LN}
\end{equation}
We shall now show that $g^{(2)}$ can be calculated explicitly
%To do so, we will apply
by a variant of the method shown in ref. \cite{Kl}.
Specifying the identity 8.530.2 of ref. \cite{GrRy} to
Hankel solutions $Z_n= H_n^{(1)} \equiv H_n$, and choosing the
$\nu$ parameter equal to zero, we obtain the summation theorem
\beq \ds H_0( m \, R(\rho, r, \varphi) ) =
\sum_{n=-\infty}^{\infty} J_n(m \rho) \, H_n( m r ) \, e^{i n
\varphi}, \hspace*{0.5cm} \ds R(\rho, r, \varphi)\equiv \sqrt{
\rho^2 + r^2 -2\rho r \cos\varphi } . \eeq Performing the change
$m \to im$, and selecting the special case $\rho=r$, it becomes
\beq \ds K_0( m \, R(r, \varphi) ) = \sum_{n=-\infty}^{\infty}
I_n(m r) \, K_n( m r ) \, e^{i n \varphi}, \hspace*{0.5cm} \ds
R(r, \varphi)\equiv  r\sqrt{ 2 (1-\cos\varphi) } =2 r \,
\left\vert \sin\left( \varphi/2 \right) \right\vert .
\label{sumtheor} \eeq Differentiating with respect to $m$, using
the property $\ds K_0'(z)= - K_1(z)$ together with the fact that
$K_n=K_{-n}$, and setting $m=1$ afterwards, we have
\begin{equation} -R(r,
\varphi) \, K_1( R(r, \varphi) ) = \sum_{n=-\infty}^{\infty} r \,
( I_n \, K_n )'(r) \, e^{i n \varphi} .
\end{equation}
Recalling the orthogonality of the
imaginary exponential functions,
we arrive at
\begin{equation}
{1 \over 2\pi}\int_0^{2\pi} d\varphi \, \left[ R(r,\varphi) \, K_1(
R(r,\varphi) ) \right]^2 = \sum_{n=-\infty}^{\infty} \left[ r (
I_n \, K_n )'(r) \right]^2 \equiv F(r) ,\label{firstsum}
\end{equation}
In order to proceed, we
rename $r$ into
$a y$, and do the variable change
$\ds u\equiv\left\vert \sin\left(\varphi/2 \right) \right\vert$.
After differentiation with respect to $y$,
we realize that the sum in \req{E2LN} is given by
\begin{equation}
g^{(2)}(y)=\frac{\hbar c \xi^2 }{4\pi }\sum_{n=-\infty}^{\infty }\,
y^2 \partial_y
\left[ y{{\partial }_y}(I_n(a y)K_n(a y))  \right]^2
=\frac{2 \hbar c \xi^2 }{\pi^2}\int _{0}^{1} du
\, \frac{u^2}{\sqrt{1-u^2}} y^2{{\partial }_y}(a y K_1(2a y u))^2 ,
\label{gdensity}
\end{equation}
Thus, we have turned the problem of calculating an infinite
angular-momentum summation into the evaluation of a definite
integral of a transcendental function. Next, by writing the
product of Bessel functions appearing in Eq.\req{gdensity} in
terms of the Meijer $G$ function \cite{Bateman},
 $g^{(2)}$ can be evaluated explicitly. (See Appendix for the
definition and some simple properties of this function.) First, we write
\begin{equation}
u^2 a^2 y^2 K_1^2 (2a y u)=u^2a^2y^2\frac{\sqrt{\pi}}{2}
G^{30}_{13}\left(\frac{1}{2};-1,0,1;4a^2y^2u^2\right)=\frac{\sqrt{\pi}}{8}
G^{30}_{13}\left(\frac{3}{2};0,1,2;4a^2y^2u^2\right)\label{product}
\end{equation}
where, in the last step, we have made use of the identity \req{id1}.
Differentiating \req{product} with respect to $y$ and substituting
in \req{gdensity} we obtain
\begin{equation}
g^{(2)}(y)=-\frac{2 \hbar c\xi^2}{\pi^{3/2}}\int_{0}^{1} du \,
\frac{y^3 a^2u^2}{\sqrt{1-u^2}}
G^{30}_{13}\left(\frac{1}{2};0,0,1;4a^2y^2u^2\right) .
\end{equation}
Using eq.$5.5.2(5)$ in \cite{Bateman}, and some straightforward
manipulation one gets:
\begin{equation}
g^{(2)}(y)= -\frac{\hbar c{\xi}^2}{8 \pi a} \,
G^{31}_{24}\left(1,2;\frac{3}{2},\frac{3}{2},\frac{5}{2},\frac{1}{2};4a^2y^2\right)
. \label{fdensity}
\end{equation}

\subsection{Calculation of the energy to order $\xi^2.$}

We now turn to the question of deriving the Casimir energy. There
are two possibilities:
\\ \ni{\bf 1.} To perform the $u$ integration in \req{gdensity} after
integrating over the $y$ variable. Unfortunately, this turns out
to be divergent. Considering the integral
\begin{equation}
%\frac{1}{\pi }
{2\over \pi^2}
\int _{0}^{\infty }{y^2} dy\, \int _{0}^{\frac{\pi}{2}}
du \,
{{\partial }_y}{ \left[ y \sin u{K_1}(2y \sin u) \right]^2 }
\end{equation}
and interchanging the order of integration, one arrives at
\begin{equation}
\int_{0}^{\infty} dy \, y^2{\partial}_y \left[ y \sin u \, K_1(2y \sin u) \right]^2 =
\frac{-1}{12 \, \sin^2 u}.
\end{equation}
If we now try to do the $u$-integration, the integral diverges.
This shows that some further regularization is called for.
In fact, in sec. \ref{seczf} we will go through the same sort of
calculation, but with the advantage of having applied zeta
function regularization from the
%very beginning.
outset.

\ni{\bf 2.} Direct integration of $g^{(2)}(y)$. We use
the following identity from ref. \cite{Bateman} (Vol 1, page 215).
\begin{equation}
\begin{array}{c}
\ds \int_{0}^{\infty } dy \, y^{-a}{K_{\nu }}\big(2\sqrt{y}\big) \,
 G^{mn}_{pq}\left(a_1 ,\dots, a_p;b_1 ,\dots, b_q;x y \right) \\ \ds
=\frac{1}{2} G^{m,n+2}_{p+2,q}\left(
a-\frac{\nu}{2},a+\frac{\nu}{2},a_1,\dots, a_p;b_1 ,\dots, b_q;x
\right)
\end{array}
\label{bateint}
\end{equation}
In order to take advantage of this formula, we note that
$\ds K_{\frac{1}{2}}(y)=\sqrt{\frac{2}{\pi y}} \, e^{-y}.$
Changing to a variable  $t=\frac{{\sigma_r}^2y^2}{4}$ and
inserting $K_{\frac{1}{2}}$,
we can cast the energy per unit-length into the form
\begin{equation}
{\cal E}^{(2)}_C \, \xi^2
=\int_{0}^{\infty}g^{(2)}(y)dy=-\lim_{\sigma_r\rightarrow 0}
\frac{\hbar c{\xi}^2}{2 a^2\pi\sqrt{\pi}\sigma_r} \, \int _{0}^{\infty}
dt \, t^{-\frac{1}{4}} K_{\frac{1}{2}}
\left(2\sqrt{t}\right)G^{31}_{24}
\left(1,2;\frac{3}{2},\frac{3}{2},\frac{5}{2},
\frac{1}{2};\frac{16t^2}{{\sigma_r}^2}\right).
\end{equation}
Although we have inserted $K_{\frac{1}{2}}$, as a mere
technicality to help calculate the energy, one can think of using
it as an {\it exponential} regulator\footnote{Actually, it is not
difficult to show that applying an exponential regulator in the
form $e^{-\sigma_r\omega}$ to \req{contoursum}, before carrying the
$k_z$ integration, yields the same result as the one we derive.}
(see ref.\cite{Kl}). However the convergence of the integral shows
that the density we have derived is already regularized in some
sense. One can now use \req{bateint} to get:
\begin{equation}
{\cal E_C}^{(2)} = -{\hbar c \over 4 a^2\pi\sqrt{\pi}\sigma_r}
G^{33}_{44}\left(0,\frac{1}{2},1,2;\frac{3}{2},\frac{3}{2},\frac{5}{2},
\frac{1}{2};\frac{16}{{\sigma_r}^2}\right) .
\end{equation}
In order to check the asymptotics as $\sigma_r\rightarrow 0$, we
use the property \req{Gproperty}, together with the asymptotics
(again from ref.\cite{Bateman}) $\ds G(x)={\cal O}(|x{|^{\beta }})
\mbox{ as $x\to 0$},$ for $p\leq q$, and $\beta =\mbox{max}\{
\mbox{Re} \, b_h \}$ for $h=1,\dots, m$. In our case we simply
have
\begin{equation}
{\cal E_C}^{(2)} \propto \lim_{\sigma_r\rightarrow 0}
\frac{1}{\sigma_r}{\cal O}(|\sigma_r^2|)=
\lim_{\sigma_r\rightarrow 0}{\cal O}(\sigma_r)=0 .
\label{Evanish}
\end{equation}
Thus, the $\xi^2$-term is shown to vanish, confirming the
conclusions of refs. \cite{MNN}, \cite{LNB} and \cite{NP}, without
recourse to numerical evaluations.

\section{Complete zeta function regularization \label{seczf}}
In this section we will take a different approach, based on
the application of the complete zeta function method (see e.g.
refs.\cite{LRap,GR}) to the initial mode sum \req{decom}.
The use of zeta functions for regularizing such sort of sums
dates from the time of refs.\cite{BM}.
In the version we shall now apply,
the regularized value of the Casimir energy
per unit-length is
\beq
{\cal E}_C =
\lim_{s \to -1}{1 \over 2} \, \hbar c \, \zeta_{\Omega (D=3)}(s)
\label{presc}
\eeq
where the zeta function
$\zeta_{\Omega (D=3)}$
for the whole set of $\omega$-modes in the three-dimensional problem
---say $\Omega$--- is given by
\beq
\zeta_{\Omega (D=3)}(s)=
\sum_{n,m} \,
\int_{-\infty}^{\infty} {d k_z\over 2\pi} \,
\left( \omega_{n,m,k_z} \over c \right)^{-s} .
\label{zetaOmD3}
\eeq
First, one assumes that $s$ is large enough for this function to make sense,
with the final aim of setting $s=-1$ at the end
(usually, one introduces in \req{presc} an arbitrary mass scale,
but,
in this problem, it turns out to be unnecessary).
Taking into account \req{decom}, we write
\beq
\zeta_{\Omega (D=3)}(s) = \sum_{n,m} \,
\int_{-\infty}^{\infty} {d k_z\over 2\pi} \,
\left[ \lambda_{n,m}^2 + k_z^2 \right]^{-s/2}
={1 \over 2\pi} \, \mbox{B}\left( {s-1\over 2}, {1 \over 2} \right) \,
\sum_{n,m} \, \lambda_{n,m}^{-(s-1)} .
\label{zetaOmD3reg}
\eeq
Let's consider the zeta function for the
projected two-dimensional problem,i.e., for the $\Lambda$ eigenmode set:
\beq
\zeta_{\Lambda (D=2)}(\sigma)= \sum_{n,m} \, \lambda_{n,m}^{-\sigma}
=\sum_{n=0}^{\infty} d_n \zeta_n(\sigma) ,
\hspace*{1cm}
\left\{
\begin{array}{lll}
d_0= 1,
&  \\
d_n= 2,
&
\mbox{ for $n\ge 1$},
\end{array}
\right.
\label{defzD2}
\eeq
%(this $\sigma$ has no immediate relation to the $\sigma_r$ of sec. \ref{secden}),
where
$\zeta_n(\sigma)$ stands for the $n$th {\it partial-wave} zeta function
\beq
\zeta_n(\sigma)=\sum_{m=1}^{\infty} \lambda_{n,m}^{-\sigma} .
\label{defznD2}
\eeq
Bearing this in mind, we put \req{zetaOmD3reg} as
\beq
\begin{array}{lll}
\ds\zeta_{\Omega (D=3)}(s)&=&\ds{1 \over 2\pi} \,
\mbox{B}\left( {s-1\over 2}, {1 \over 2} \right) \,
\zeta_{\Lambda (D=2)}(s-1) \\
&=&\ds{1 \over 2\pi} \,
\left[
{ \zeta_{\Lambda (D=2)}(-2) \over s+1 }
+\left( \ln(2)-{1\over 2} \right) \, \zeta_{\Lambda (D=2)}(-2)
+\zeta_{\Lambda (D=2)}'(-2)
+{\cal O}(s+1)
\right] ,
\end{array}
\label{zD3sexp}
\eeq
where an expansion around $s=-1$ has taken place.
This was the method  applied in ref.\cite{GR}.

Eqs. \req{defzD2}, \req{defznD2} hold for $\rep \sigma > 1$, but they will have to be analytically
continued to the neighbourhood of $\sigma=-2$ (note that $\sigma= s-1$).
Such an analytic continuation
is carried out by the contour integration method of refs.\cite{ELRLRjpa,LRap}.
To begin with, one takes
\beq
a^{-\sigma} \, \zeta_n(\sigma)=
{\sigma \over 2\pi i} \int_C du \, u^{-\sigma-1} \, \ln\left[ f_n(u) \right] ,
\mbox{ for $\rep \sigma > 1$ } ,
\label{znsg1}
\eeq
where $f_n(u)\equiv f_n(k_z, \omega, a)$ with $u\equiv \lambda a$, and $C$ is
a circuit in the complex $u$-plane enclosing all the positive zeros of
$f_n(u)$. In the desired limit this contour
will be semicircular, with the straight parts along the imaginary axis, and
adequately avoiding the origin.
The first step (see e.g.\cite{ELRLRjpa})
is to examine the
asymptotic behaviour $f_{n, \text{as}}(u)$ of $f_n(u)$ for $|u| \to \infty$.
If $f_{n, \text{as}}(u)$ has no roots inside of $C$, we
leave the eq.\req{znsg1} unchanged
by setting
\beq
a^{-\sigma} \, \zeta_n(\sigma)={\sigma \over 2\pi i} \int_C du \, u^{-\sigma-1} \,
\ln\left[ f_n(u) \over f_{n, \text{as}}(u) \right] .
\label{znsg1-2}
\eeq
Going back to \req{defcalPW},
we observe that, for large $x$,
${\cal P}_n^2(x)={\cal O}\left( x^{-4} \right)$.
Then,
one can write
\beq
f_n(u) = \ds f_{n, \text{as}}(u) \,
\left[
1+\xi^2{\pi^2 u^2\over 4}
{\cal P}_n^2(u)
\right],
\hspace*{1cm}
f_{n, \text{as}}(u) =
-{ (\varepsilon_1+\varepsilon_2)^2 \over \varepsilon_1 \, \varepsilon_2  } \,
{c^{-2} \over \pi^2 a^4} \, u^4.
\label{fnas}
\eeq
Therefore, eq.\req{znsg1-2} translates into
\beq
a^{-\sigma} \zeta_n(\sigma)={\sigma \over 2\pi i} \int_C du \, u^{-\sigma-1} \,
\ln\left[ 1+ \xi^2 {\pi^2\over 4} \, u^2 {\cal P}_n^2(u) \right] .
\label{znsg1-3}
\eeq
After realizing
that only the vertical parts of $C$  ---where $u=e^{\pm i\pi/2}y$---
are actually contributing to the integration,
eq.\req{znsg1-3} yields
\beq
\begin{array}{c}
\ds \zeta_n(\sigma)=
a^{\sigma} \, {\sigma \over \pi} \sin\left( \pi \sigma \over 2 \right) \,
\int_0^{\infty} dy \, y^{-\sigma-1} \,
\ln\left\{ 1 -\xi^2\left[ y \, \left( I_n \, K_n \right)'(y) \right]^2 \right\} ,
\mbox{ for $-1 < \rep \sigma < 0$},
\end{array}
\label{znsg1-4}
\eeq
where we have
%taken into account that
used
$\ds {\cal P}_n^2(\pm i y)=
{4 \over \pi^2}\left[ \left( I_n \, K_n \right)'(y) \right]^2$,
being $I_n$, $K_n$ the corresponding modified Bessel functions
(note that this $y$ is dimensionless).
All this has validity near $\sigma =-1$, but we still need some further
work in order to reach the neighbourhood of $\sigma =-2$.

\subsection{\bf Calculation to the order of $\xi^2$}
Let $\ds {\cal E}_C= \sum_{p \geq 1} {\cal E}_C^{(2p)} \, \xi^{2p},$
and analogously for the involved zeta functions. Then,
\beq
\begin{array}{rcl}
\zeta_n(\sigma)&=&\ds -a^{\sigma} \, {\sigma \over \pi} \sin\left( \pi \sigma \over 2 \right)
\,
\sum_{p\geq 1}{1 \over p} \xi^{2p} \, A_n^{(2p)}(\sigma), \\
\ds A^{(2p)}_n(\sigma)&=&\ds \int_0^{\infty} dy \, y^{-\sigma-1} \,
\left[ y \, (I_n \, K_n)'(y) \right]^{2p},
\mbox{\hspace*{5mm} for $-1 < \rep \sigma < 0$}.
\end{array}
\label{defAn}
\eeq
Note that
we are commuting a $\xi$-expansion
with a process of analytic extension which sidesteps $\sigma$-poles (i.e., $s$-poles).
Yet, since the $\xi$-dependence has no problematic traits,
this should be correct,
and we write
\beq
\begin{array}{rcl}
\ds \zeta_{\Lambda(D=2)}(\sigma)&=&\ds\sum_{n=0}^{\infty} d_n \, \zeta_n(\sigma) \
= \sum_{p \geq 1} \zeta_{\Lambda(D=2)}^{(2p)}(\sigma) \, \xi^{2p}, \\
\ds \zeta_{\Lambda(D=2)}^{(2p)}(\sigma)
&=&\ds -{1 \over p} \, a^{\sigma} \, {\sigma \over \pi}\sin\left( \pi \sigma \over 2 \right) \,
\sum_{n= 0}^{\infty} d_n \, A^{(2p)}_n(\sigma) .
\end{array}
\label{defzD22p}
\eeq
If we just want to keep the terms $\sim\xi^2$ in ${\cal E}_C$,
it will be enough to maintain
the $p=1$ contribution, which can be rewritten in the way
\beq
\ds \zeta_{\Lambda(D=2)}^{(2)}(\sigma)=\ds
-a^{\sigma} \, {\sigma \over \pi}\sin\left( \pi \sigma \over 2 \right) \,
\int_0^{\infty} dy \, y^{-\sigma-1} \, F(y) ,
\hspace*{5mm}
F(y)\equiv\ds \sum_{n=-\infty}^{\infty} \left[ y (I_n K_n)'(y) \right]^2 ,
\label{defFy}
\eeq
where we have also taken into account
\req{defzD2} and
the fact that $\ds \zeta_{-n}(\sigma)= \zeta_n(\sigma)$.
An integral representation of the $F(y)$ is already available
in eq.\req{firstsum}. From
there,
we proceed as in the derivation of eq.\req{gdensity}, i.e.,
we do the variable change $u \equiv |\sin(\varphi/2)|$
and find
\begin{equation}
F(y)={8 \, y^2 \over \pi} \int_0^1 {du \over \sqrt{1-u^2}} \, u^2\,
K_1^2(2 u y) . \label{Frint}
\end{equation}
With this, we go back to eq. \req{defFy} and focus on the integral
\begin{equation}
{\cal F}(\sigma)\equiv \int_0^{\infty} dy \, y^{-\sigma-1} \, F(y)
= {8 \over \pi} \int_0^1 {du \over \sqrt{1-u^2}} \, u^2 \,
\int_0^{\infty} dy \, y^{-\sigma+1} \, K_1^2(2 u y) .
\label{defcalF}
\end{equation}
The $y$-integration is
%straightforwardly
evaluated with the help
of formula 6.576.4 in ref.\cite{GrRy}.
Then, the remaining $u$-integral is immediate using forumla 3.251.1 in
the same book.
As a result,
\beq
{\cal F}(\sigma)= {1 \over 2 \, \sqrt{\pi}} \,
{\Gamma\left( 4-\sigma \over 2 \right) \Gamma^2\left( 2-\sigma \over 2 \right)
\Gamma\left( -{\sigma \over 2} \right)
\over \Gamma(2-\sigma) \Gamma\left( \sigma+2\over 2 \right) } ,
\eeq
which has
a zero of order one at $\sigma= -2$
by virtue of the singularity of $\Gamma\left( \sigma+2\over 2 \right)$.
Putting it into eq.\req{defFy} and expanding near $\sigma=-2$,
we find
\beq
\ds\zeta_{\Lambda(D=2)}^{(2)}(\sigma)
=\ds -a^{\sigma} \, {\sigma \over \pi}\sin\left( \pi \sigma \over 2 \right) \,
{\cal F}(\sigma)
=\ds
{1 \over a^2 }
\left[
{1 \over 6} (\sigma+2)^2 + {\cal O}\left( (\sigma+2)^3 \right)
\right] ,
\eeq
which provides the desired analytic extension
to $\rep \sigma = -2$.
The crucial point is that it has a zero of order two at $\sigma=-2$
and, therefore,
$\zeta_{\Lambda(D=2)}^{(2)}(-2)=0$ and
${\zeta_{\Lambda(D=2)}^{(2)}}'(-2)=0$ .
This, together with eqs. \req{presc} and \req{zD3sexp},
leads to
${\cal E}_C= 0 + {\cal O}\left( \xi^4 \right)$, i.e.,
\beq
{\cal E}_C^{(2)} = 0,
\label{E2res}
\eeq
which was numerically found in ref.\cite{MNN}
\footnote{Incidentally, the comments made by one the
authors of the present letter led to the correct numerical figure in ref.\cite{MNN}.}
(see also \cite{LNB,NP}).

\subsection{\bf Higher-order corrections in $\xi^2$}
In order to know new corrections in $\xi^2$, one has to keep the
next $\xi^2$-terms in
eqs. \req{defAn}, \req{defzD22p}.
However,
unlike $A^{(2)}_n(\sigma)$,
the $A^{(2p)}_n(\sigma)$ integrals with $p \geq 2$
are already finite at $\sigma=-2$,
because
$\ds\left[ y (I_n K_n)'(y) \right]^{2p} \sim {1 \over (2 y)^{2p} }$
as $y\to\infty$.
Thus, the restriction to $\rep \sigma > -1$ in \req{defAn}
is caused only by the presence of $p=1$, while,
for $p \geq 2$, it suffices to
numerically evaluate all the necessary $A^{(2p)}_n(\sigma)$'s at
$\sigma=-2$,
i.e.,
\beq
{\zeta_{\Lambda(D=2)}^{(2p)}}'(-2) = -{1 \over p \, a^2} \,
\sum_{n=0}^{\infty} d_n \, A^{(2p)}_n(-2).
\eeq
{\it A posteriori}, we have verified that
each term decreases quickly enough with $n$ and
that the $n$-summation has a numerically acceptable behaviour.
Then, the finiteness of each of these sums confirms that
$\zeta_{\Lambda(D=2)}^{(2p)}(-2)= 0$, and
the $\xi^{2p}$-contributions to ${\cal E}_C$ are
\beq
\begin{array}{rcl}
\ds {\cal E}_C^{(2p)} \, \xi^{2p}&=&\ds\hbar c \, {1 \over 2} \,
{\zeta_{\Omega(D=3)}^{(2p)}}'(-1) \, \xi^{2p}
= \ds\hbar c \, {1 \over 4\pi} \,
{\zeta_{\Lambda(D=2)}^{(2p)}}'(-2) \, \xi^{2p} \\
&=&\ds -\hbar c \, {\xi^{2p} \over 4 \pi \, p \, a^2} \,
\sum_{n=0}^{\infty} d_n \, A^{(2p)}_n(-2) ,
\hspace*{3cm}\mbox{for $p \geq 2$,}
\end{array}
\label{E2psumA2p}
\eeq
where the meaning of $\zeta_{\Omega(D=3)}^{(2p)}(s)$ is obvious.
When $p=2$,
including all the $n$-values up to $n_{\text{max}} \sim 120$,
we have found
$\ds\sum_{n\geq 0} d_n \, A^{(4)}_n(-2) \simeq 0.19108$.
This and formula \req{E2psumA2p} yield
\beq
{\cal E}_C^{(4)} \, \xi^4=
-0.0076028 \, { \hbar c \over a^2} \, \xi^4 ,
\label{E4res}
\eeq
in agreement with ref. \cite{NP}. As remarked there, the
negative sign means that the involved Casimir forces are attractive.
Physical implications concerning
the flux tube model for confinement have been discussed in that work.

%Apart from $\xi^4$,
%this type of procedure can be applied to higher orders in $\xi^2$.
In fact, the higher $p$, the fewer terms are needed in the $n$-series for
obtaining reliable figures. For $p > 2$ we have found many
contributions, but we just list the first ones:
\begin{center}
\begin{tabular}{c||c|c|c|c|c|c}
$p$&3&4&5&6&7&$\dots$ \\ \hline
${\cal E}_C^{(2p)} a^2 /(\hbar c)$
&$-0.0022637$
&$-0.0010807$
&$-0.0006202$
&$-0.0003972$
&$-0.0002737$
&$\dots$ \\
\end{tabular}
\end{center}

As argued in refs.\cite{MNN} or \cite{LNB}, the special value $\xi^2=1$ should
reproduce the perfectly-conducting case
${\cal E}_{\text{C (p.c.)}} \, a^2/(\hbar c)= -0.01356\dots $ \cite{RM,GR}.
Taking all the contributions up to $p=7$,
we obtain
${\cal E}_C \, a^2/(\hbar c)\simeq -0.01224$,
with a 10\percent relative error.
This is not too surprising, as the $\xi^2$-expansion
comes from a logarithimc series, and a slow numerical convergence at $\xi^2=1$
is expectable.
Including all the terms up to $p=200$
we have found
${\cal E}_C \, a^2/(\hbar c)\simeq -0.01354$,
with a 0.15\percent relative error.

\section{Conclusions}
The ultimate consequences of any result about Casimir effect are not
easy to foresee, as the domain of applicability of this concept has been
expanding beyond what could be considered `traditional' areas
of field theory. For instance, we have recent examples of these
ideas in spacetime evolution and quantum cosmology \cite{AB}.
Proposals haven even been made about
possible ways of extracting work from the vacuum energy \cite{Fo}.

In the present letter we have confirmed the expectation that the
$\xi^2$-contribution to the Casimir energy for a dilute-dielectric
cylinder, infinitely long, and under the condition of
light-velocity conservation, would have to vanish. Numerically
speaking, this had been noticed with very high accuracy in several
articles, starting with ref.\cite{MNN}, but in the present letter
we have been able to derive it as an {\it exact} result
(eqs.\req{Evanish} and \req{E2res}). Another new aspect lies in
applying the method developed in \cite{Kl}, i.e., the use of
summation theorems for infinite series of Bessel functions, which
has spared us the handling of Debye expansions (see also the
application of this method in ref.\cite{Ma} and \cite{Lambiase},
but this time in connection with the problem of
refs.\cite{Mdielball}).

Moreover, by a numerical evaluation, and within the complete zeta function
regularization framework,
we have reobtained the $\xi^4$-contribution calculated in ref.\cite{NP},
(our eq. \req{E4res}), which is negative.
This constitutes the first deviation from zero and shows that, although at a
higher order, the Casimir energy of this system would tend to contract the cylinder.
Even higher corrections in $\xi^2$ have also been found (table in sec. \ref{seczf}).

Our spectral zeta function has been constructed
like in ref.\cite{GR}.
Other variants of the zeta function procedure, which differ from ours at some
particular steps, are also in circulation (e.g. ref.\cite{NPo} or \cite{LNB}).
We
regard them as slightly different formulations of one common underlying principle.
In particular,
ref.\cite{NPo} illustrates the advantages of dealing with the total zeta
function as a whole object, rather than a series of partial-wave zeta functions.

\appendix\section{Appendix: The Meijer $G$ function}
Here we state some facts about the Meijer $G$ function, which
is defined by the integral
\begin{equation}
G_{pq}^{mn}(a_{\text{list}},b_{\text{list}},x)=\frac{1}{2\pi i}
\int_{L} ds \, \frac{\prod_{j=1}^{m}\Gamma ({b_j}-s)\prod
_{j=1}^{n}\Gamma (1-{a_j}+s)}{\prod _{j=m+1}^{q}\Gamma
(1-{b_j}+s)\prod _{j=n+1}^{p}\Gamma ({a_j}-s)}{x^s}.
\label{IntRepG}
\end{equation}
The different integration paths $L$ can be found, for example, in
\cite{Bateman}, as most of the other properties we use. By simple
variable changes one may prove numerous identities such as:
\begin{equation}
x^n G_{pq}^{mn}(a_{\text{list}},b_{\text{list}},x)=
G_{pq}^{mn}(a_{\text{list}}+n,b_{\text{list}}+n,x) \label{id1}
\end{equation}
and
\begin{equation}
G_{pq}^{mn}\big({a_{\text{list}}},{b_{\text{list}}},\frac{1}{x}\big)=
G_{qp}^{mn}(1-{b_{\text{list}}},1-{a_{\text{list}}},x)\label{Gproperty}
\end{equation}
which we have used in secs. $2.1, 2.2$.

\vspace*{1cm}
\ni{\Large\bf Acknowledgements}

I.K. wishes to thank J. Feinberg, A. Mann and M. Revzen for their
remarks. A.R. thanks K.A. Milton and V.V. Nesterenko for discussions.

\end{document}